\documentclass{iopart}
\usepackage{iopams}
\begin{document}
\title[Ho\v{r}ava-Lifshitz gravity  and the mixmaster universe]{\textbf{Ho\v{r}ava-Lifshitz gravity inspired Bianchi-II cosmology and the mixmaster universe}}
\author{Leonardo Giani$^1$, Alexander Y ~Kamenshchik$^{1,2}$} 
\address{$^1$Dipartimento di Fisica e Astronomia and INFN, \small Via Irnerio 46, 40126, Bologna,
Italy,\\
$^2$L.D. Landau Institute for Theoretical Physics of the Russian
Academy of Sciences, Kosygin str. 2, 119334, Moscow, Russia}
\eads{\mailto{leonardo.giani@studio.unibo.it}, \mailto{kamenshchik@bo.infn.it}}

\begin{abstract}
We study  different aspects of the Ho\v{r}ava-Lifshitz inspired Bianchi-II cosmology and its relations with the mixmaster universe model. First,  
we present exact solutions for a toy model, where only the cubic in spatial curvature terms are present in the action; then we briefly discuss some exotic singularities, which can appear in this toy model.  
We study also the toy model where only the quadratic in spatial curvature terms are present in the action.
We establish relations between our results and those obtained by using the Hamiltonian formalism. Finally, we apply the results obtained by studying Bianchi-II cosmology to describe the evolution of the mixmaster universe in terms of the Belinsky-Khalatnikov-Lifshitz  formalism.
Generally, our analysis gives some arguments in favour of the existence of the oscillatory approach to the singularity in a universe governed by the Ho\v{r}ava-Lifshitz type gravity.
\end{abstract}
\submitto{\CQG}
\maketitle

\section{Introduction}

It is generally recognised that the complete theory of elementary particles and fundamental interactions should include quantum gravity. However, the quantum gravity theory is non-renormalizable (see e.g. \cite{Weinberg}). The main obstacle against perturbative renormalizability of the general relativity in $3+1$ dimensions is the fact that the gravitational coupling constant (the Newton constant) is dimensionful with a negative dimension $[G_N]=-2$ in mass units. The graviton propagator as all the propagators in quantum field theory scales with the four-momentum $k_{\mu}=(\omega,\vec{k})$ as 
\begin{equation*}
\frac{1}{k^2},
\label{Hor}
\end{equation*}
where $k=\sqrt{\omega^2-\vec{k}^2}$. When we calculate Feynman diagrams with an increasing number of loops, it is necessary to introduce again and again the counterterms with increasing degree in curvature. 

An improved ultraviolet behaviour of the theory can be obtained if some higher-order in curvature terms are added to the Lagrangian. The terms quadratic in curvature not only yield new interactions (with a dimensionless coupling), but also modify the propagator. 
Omitting the tensor structure of this propagator, one can write it as 
\begin{eqnarray}
&&\frac{1}{k^2}+\frac{1}{k^2}G_Nk^4\frac{1}{k^2}+\frac{1}{k^2}G_Nk^4\frac{1}{k^2}G_Nk^4\frac{1}{k^2}+\cdots\nonumber \\
&&=\frac{1}{k^2-G_Nk^4}.
\label{Hor1}
\end{eqnarray}
It is easy to see that at high energies the propagator is dominated by $1/k^4$ term. This cures the problem of ultraviolet divergences. However, a new problem arises: The resummed propagator (\ref{Hor1}) has two poles:
\begin{equation*}
\frac{1}{k^2-G_Nk^4}= \frac{1}{k^2}-\frac{1}{k^2-1/G_N}.
\label{Hor2}
\end{equation*}
One of these poles describes massless gravitons, while the other corresponds to ghost excitations and implies violations of unitarity.  

Some years ago P. Ho\v{r}ava has introduced a new class of gravity models \cite{Hor-Lif},
which he has called ``Quantum gravity at a Lifshitz point''. The creation of this approach was inspired by the papers by E. M. Lifshitz \cite{Lif-phase}, devoted to the theory of second-order phase transitions and of critical phenomena and published in 1941. The Ho\v{r}ava-Lifshitz gravity models exhibit scaling properties which are anisotropic between space and time. The degree of anisotropy between space and time is measured by the dynamical critical exponent $z$ such that 
\begin{equation*}
\vec{x} \rightarrow b\vec{x},\ \ t \rightarrow b^z t.
\label{z-crit}
\end{equation*}

In the approach to  quantum gravity, suggested in \cite{Hor-Lif}, there were considered actions such that scaling at short distances exhibited a strong anisotropy between space and time, with $z > 1$. This improves the short-distance behaviour of the theory. Indeed, the propagator for such gravitons is proportional to 
\begin{equation*}
\frac{1}{\omega^2-c^2\vec{k}^2-G(\vec{k}^2)^z}.
\label{prop-Lif}
\end{equation*}

At high energies the propagator is dominated by the anisotropic term $1/(\omega^2-G(\vec{k}^2)^z)$. For a suitably chosen $z$, this modification improves the short-distance behaviour and the theory becomes power-counting renormalizable. The $c^2\vec{k}^2$ term becomes important at low energies, where the theory naturally flows to $z=1$. 

Unlike in relativistic higher-derivative theories mentioned above, higher-order time derivatives are not generated, and the problem with ghost excitations and non-unitarity is resolved. In the case of the $3+1$ gravity, the renormalizability is achieved by the choice of $z=3$. 

Rather a large amount of work was connected with applications of the Ho\v{r}ava-Lifshitz gravity to cosmology (for a review see \cite{Hor-cosm}). 
The presence of different terms, depending on the spatial curvature  in the action, can provide the potentials of rather a complicated form which can give rise to such effects as the appearance of the cyclic universe \cite{Maeda} or to obtain 
the scale invariant power spectrum without inflation \cite{power}. The Bianchi-IX universe dynamics in Ho\v{r}ava-Lifshitz gravity was studied in paper \cite{Maeda1}.
The cosmology of the Bianchi-III universe was studied in paper \cite{Bianchi-III}. The Hamiltonian approach to the cosmology in the Ho\v{r}ava-Lifshitz gravity was 
developed in papers \cite{Kluson, Bakas}. The non-singular Ho\v{r}ava-Lifshitz cosmological models were studied in paper \cite{non-sing}. The phase-space analysis of the late-time asymptotical regimes in the Ho\v{r}ava-Lifshitz cosmologies was done in paper \cite{late-time}.

The present paper is motivated by the intention to understand what happens with the universe driven by  the Ho\v{r}ava-Lifshitz gravity, when the former is very close to the cosmological singularity.
The problem of the existence of the cosmological singularity at the beginning of  cosmological evolution has attracted the  attention of  people  studying general relativity for a long time \cite{Khal-Lif}.   In papers \cite{Pen-Hawk}
the impossibility  of the indefinite continuation of geodesics under certain conditions was shown.
This was interpreted as pointing to the existence of a singularity in the general
solution of the Einstein equations.
The analytical behaviour of the general solutions to the Einstein equations in the
neighbourhood of a singularity was investigated in papers \cite{BKL}.
These papers revealed the enigmatic  phenomenon of an oscillatory approach to the singularity
which has also become known as the {\it Mixmaster Universe} \cite{Misner}.
The model of a closed homogeneous, but anisotropic, universe with three degrees of freedom
(Bianchi IX cosmological model) was used to demonstrate that the universe approaches the singularity
in such a way that its contraction along two axes is accompanied by an expansion with respect to
the third axis, and the axes change their roles according to a rather peculiar law which reveals
a chaotic behaviour~\cite{BKL,chaos}.

Later the mixmaster type behaviour was studied also in multidimensional Kaluza-Klein cosmology \cite{multi} (for a recent review see e.g. \cite{multi1}). 
Especially interesting is the study of the chaotic behaviour in the superstring-inspired cosmological models \cite{superstring}, where an unexpected relation
between chaotic behavior and the hyperbolic infinite dimensional Lie algebras \cite{Kac} was established. 

Thus, it is of particular interest also to try to find out what happens with a mixmaster type universe driven by some kind of the Ho\v{r}ava-Lifshitz gravity. 
The papers \cite{Bakas,Hor-mix} were devoted to this problem.
It looks like  the question about the presence or absence of such a  stochasticity in this context  is not yet resolved and requires further studies.   

In this paper we would like to study the applications of the Ho\v{r}ava-Lifshitz gravity to the cosmology of a Bianchi-II universe, hoping that such an investigation can help to understand better the situation with a possibility of the oscillatory approach to the singularity and of the stochasticity arising in this context. 
Indeed, it is known that in general relativity the Einstein equations for the Bianchi-II universe are exactly solvable and give also asymptotic regimes, which 
arise in the Bianchi-IX universe. Thus, it makes sense to write down the corresponding solutions for some Ho\v{r}ava-Lifshitz inspired models, or in other words, in the models where some essential terms depending on the spatial curvature of the universe are explicitly taken into account. It permits us to make some observations, concerning the behaviour of the Bianchi-IX mixmaster universe. We shall use basically the BKL formalism \cite{BKL}, however, we shall confront our results with those obtained in paper \cite{Bakas}, where the Hamiltonian formalism \cite{Misner} was applied.

Let us note that in what follows we do not take into account the possible presence of matter in the universe. It is justified by the fact that in the vicinity of the cosmological singularity, the presence of dust-like or radiation-like matter does not change the behaviour of the universe for homogeneous cosmologies \cite{BKL}. It is connected with the fact that the corresponding terms in the Einstein equations in the vicinity of the singularity are small with respect to those originated from the spatial curvature or anisotropies.
The matter becomes essential when the expansion of the universe drives it far away from the singularity. Then the matter becomes responsible for the isotropization of  
 the universe. Such an effect is well studied in the literature \cite{Heck-Schuck, Khal-Kam} and the corresponding behaviour in the Ho\v{r}ava-Lifshitz gravity is not different from that in general relativity. 

The structure of the paper is as follows: in Sec. 2 we write down the explicit solution of the Einstein equations for the Bianchi-II universe; the Sec. 3 is devoted to the study of different aspects of the Ho\v{r}ava-Lifshitz inspired Bianchi-II cosmology; in subsection 3.1 we present exact solutions for a toy model, where only the cubic in spatial curvature terms are present in the action; in subsection 3.2 we briefly discuss some exotic singularities, which can appear in this toy model; 
in subsection 3.3 we write down the explicit solutions for the model, where the greatest terms are those coming from the invariants quadratic in spatial curvature.
In the fourth section we establish relations between our results and those, obtained by using the Hamiltonian formalism. In the fifth section we apply the results obtained in the section 3 to describe the evolution of the mixmaster universe in terms of the BKL formalism.  
 The last section contains some concluding remarks. Generally, our analysis gives some arguments in favour of the existence of the oscillatory approach to the singularity.   

\section{The evolution of the Bianchi-II universe in general relativity}

To begin with, let us recollect briefly what  is known about the evolution of the empty Bianchi-II universe governed by the standard Einstein equations. 
The metric of the Bianchi-II universe is \cite{Land-Lif, Ryan}
\begin{equation*}
ds^2=dt^2-a^2(t)(\omega^1)^2-b^2(t)(\omega^2)^2-c^2(t)(\omega^3)^2,
\label{Bianchi-II}
\end{equation*}
where the basis one-forms are given by 
\begin{eqnarray*}
&&\omega^1 = dy-xdz,\nonumber \\
&&\omega^2 = dz,\nonumber \\
&&\omega^3 = dx.
\label{omega}
\end{eqnarray*}
It is convenient  to introduce a new time parameter $\tau$
\begin{equation*}
dt = a(t)b(t)c(t) d\tau
\label{tau}
\end{equation*}
and to represent the scale factors as 
\begin{eqnarray*}
&&a = e^{\alpha(\tau)},\nonumber \\
&&b = e^{\beta(\tau)},\nonumber \\  
&&c = e^{\gamma(\tau)}.
\label{scale}
\end{eqnarray*}

Now, the spatial components of the Einstein equations $R_a^a=0$ look like \cite{Land-Lif}
\begin{equation}
\ddot{\alpha}=-\frac12e^{4\alpha},
\label{alpha}
\end{equation}
\begin{equation}
\ddot{\beta}=\frac12e^{4\alpha},
\label{beta}
\end{equation}
\begin{equation}
\ddot{\gamma}=\frac12e^{4\alpha},
\label{gamma}
\end{equation}
while the temporal component of the Einstein equations $R_0^0=0$ is 
\begin{equation}
\ddot{\alpha}+\ddot{\beta}+\ddot{\gamma} = 2(\dot{\alpha}\dot{\beta}+\dot{\alpha}\dot{\gamma}+\dot{\beta}\dot{\gamma}).
\label{00}
\end{equation}
Here ``dot'' means the differentiation with respect to the time parameter $\tau$. 

Multiplying the equation (\ref{alpha}) by $\dot{\alpha}$ we immediately find the first integral
\begin{equation*}
\dot{\alpha}^2=H^2-\frac14e^{4\alpha},
\label{alpha1}
\end{equation*}
where $H$ is a constant. Hence,
\begin{equation*}
\dot{\alpha}=\pm\sqrt{H^2-\frac14e^{4\alpha}}.
\label{alpha2}
\end{equation*}
This equation is also integrable and one has
\begin{equation}
\alpha(\tau) = \mp H\tau -\frac12\ln(e^{\mp4H\tau}+1),
\label{alpha3}
\end{equation}
where the additional constant of integration is absorbed by a shift of the time parameter.
The sum of Eqs. (\ref{alpha}) and (\ref{beta}) gives
\begin{equation*}
\ddot{\alpha}+\ddot{\beta}=0
\label{beta1}
\end{equation*}
and 
\begin{equation*}
\dot{\alpha}+\dot{\beta}=B,
\label{beta2}
\end{equation*}
where $B$ is an integration constant. 
Hence
\begin{equation*}
\beta(\tau) = B\tau+B_0-\alpha(\tau).
\label{beta3}
\end{equation*}
In what follows we shall omit the constant $B_0$ because its role is reduced to the constant rescaling of the factor $b(t)$.
Quite analogously we have 
\begin{equation*}
\gamma(\tau) = C\tau+C_0-\alpha(\tau).
\label{gamma3}
\end{equation*}
Now, substituting the obtained solutions into the Einstein $00$ equation, we find the following relation between the constants 
$H, B$ and $C$:
\begin{equation}
BC=H^2.
\label{constraint}
\end{equation}

Now, we can choose the sign plus in Eq. (\ref{alpha3}). In this case when $\tau \rightarrow -\infty$ the logarithmic scale factors 
 behave as 
 \begin{eqnarray*}
 &&\alpha(\tau) = H\tau,\nonumber \\
 &&\beta(\tau) = (B-H)\tau,\nonumber \\
 &&\gamma(\tau) = (C-H)\tau.
 \label{min-inf}
 \end{eqnarray*}
 The spatial volume of the universe behaves as
 \begin{equation*}
 abc=\exp(\alpha+\beta+\gamma)=\exp[(B+C-H)\tau].
 \label{min-inf1}
 \end{equation*}
 When $\tau \rightarrow \infty$, we have 
 \begin{eqnarray*}
 &&\alpha(\tau)=-H\tau,\nonumber \\
 &&\beta(\tau)=(B+H)\tau,\nonumber \\
 &&\gamma(\tau)=(C+H)\tau.
 \label{plus-inf}
 \end{eqnarray*}
The spatial volume of the universe is 
 \begin{equation*}
 abc=\exp(\alpha+\beta+\gamma)=\exp[(B+C+H)\tau].
 \label{plus-info}
 \end{equation*}
Choosing all three constants $H, A$ and $B$ positive we see that at the moment $\tau \rightarrow -\infty$ the universe is closer to the singularity, than at the moment $\tau \rightarrow \infty$. Making the transition to the cosmic time $t$, one can see that at the moment, corresponding to $\tau \rightarrow \infty$, the scale factors behave as 
\begin{eqnarray}
&&a(t) \sim t^{-\frac{H}{B+C+H}}=t^{p_1},\nonumber \\
&&b(t) \sim t^{\frac{B+H}{B+C+H}}=t^{p_2},\nonumber \\
&&c(t) \sim t^{\frac{C+H}{B+C+H}}=t^{p_3}.
\label{plus-inf2}
\end{eqnarray}
It is easy to check that the exponents in Eqs. (\ref{plus-inf2}) are the Kasner indices, satisfying the equations 
\begin{eqnarray}
&&p_1+p_2+p_3=1,\nonumber \\
&&p_1^2+p_2^2+p_3^2=1.
\label{Kasner}
\end{eqnarray}
Introducing the Khalatnikov-Lifshitz parameter $u=H/B$ \cite{Khal-Lif} and taking into account the relation (\ref{constraint}) we see that the
Kasner indices have the standard parametrisation
\begin{eqnarray}
&&p_1=-\frac{u}{1+u+u^2},\nonumber \\
&&p_2=\frac{1+u}{1+u+u^2},\nonumber \\
&&p_3=\frac{u(1+u)}{1+u+u^2}.
\label{Kasner1}
\end{eqnarray}
When the universe moves close to the singularity $\tau \rightarrow -\infty$ we have 
\begin{eqnarray}
&&a(t) \sim t^{\frac{H}{B+C-H}}=t^{p'_1},\nonumber \\
&&b(t) \sim t^{\frac{B-H}{B+C-H}}=t^{p'_2},\nonumber \\
&&c(t) \sim t^{\frac{C-H}{B+C-H}}=t^{p'_3}.
\label{min-inf5}
\end{eqnarray}
Here the Kasner indices are given by 
\begin{eqnarray}
&&p'_1=\frac{u}{1-u+u^2}=p_2(u-1),\nonumber \\
&&p'_2=\frac{1-u}{1-u+u^2}=p_1(u-1),\nonumber \\
&&p'_3=\frac{u(u-1)}{1-u+u^2}=p_3(u-1).
\label{Kasner2}
\end{eqnarray}
Thus, we have reproduced the known law of the transition from one Kasner epoch to another \cite{BKL, Land-Lif}.

Let us notice that above we have considered the situation when at $\tau \rightarrow \infty$ the Kasner index corresponding to the scale factor $a$, which stays in the numerator of the  potential term, has a sign different from that of the two other scale factors 
$b$ and $c$, which stay in the denominator of the potential term. Such a choice was motivated by the fact that in a Bianchi-IX universe, where all three types of the contributions to the potential $\frac{a^2}{b^2c^2}, \frac{b^2}{a^2c^2}$ and $\frac{c^2}{a^2b^2}$ are present, one takes into account only the dominant term \cite{BKL, Land-Lif}. In a Bianchi - II universe there is only one potential term $\frac{a^2}{b^2c^2}$ and the exact solution is valid independently on the relation between the Kasner indices of the scale factors in the asymptotic regime. Thus,   we shall consider also the situation when at $\tau \rightarrow \infty$ the Kasner index of the 
scale factor $a$ is positive while the Kasner index of the scale factor $b$ is negative.  It can be realised if the integration constant $H$ is negative: $H = -|H|$, while the constant $B$ is positive and $B < |H|$. In this case it is convenient to introduce the parameter $u$ as follows:
\begin{equation}
u=\frac{|H|}{B}-1,
\label{u-new}
\end{equation}
 and 
 \begin{eqnarray}
&&a(t) \sim t^{\frac{|H|}{B+C-|H|}}=t^{p_2},\nonumber \\
&&b(t) \sim t^{\frac{B-|H|}{B+C-|H|}}=t^{p_1},\nonumber \\
&&c(t) \sim t^{\frac{C-|H|}{B+C-|H|}}=t^{p_3},
\label{plus-inf4}
\end{eqnarray}
where the Kasner indices are given again by Eq. (\ref{Kasner1}). When $\tau \rightarrow -\infty$, i.e. when the universe is closer to the singularity, we have 
\begin{eqnarray}
&&a(t) \sim t^{\frac{-|H|}{B+C+|H|}}=t^{p'_2},\nonumber \\
&&b(t) \sim t^{\frac{B+|H|}{B+C+|H|}}=t^{p'_1},\nonumber \\
&&c(t) \sim t^{\frac{C+|H|}{B+C+|H|}}=t^{p'_3}.
\label{min-inf6}
\end{eqnarray}
However, now the relations between the values of the Kasner indices at the asymptotic regimes is different and are given by 
\begin{eqnarray}
&&p'_1=\frac{(u+1)+1}{(u+1)^2+(u+1)+1}=p_2(u+1),\nonumber \\
&&p'_2=\frac{-(u+1)}{(u+1)^2+(u+1)+1}=p_1(u+1),\nonumber \\
&&p'_3=\frac{(u+1)[(u+1)+1]}{(u+1)^2+(u+1)+1}=p_3(u+1).
\label{Kasner-new}
\end{eqnarray}
Thus, we see that in this case, instead of the standard shift of the Khalatnikov-Lifshitz parameter $ u \rightarrow u-1$, one has 
$u \rightarrow u+1$.  

\section{Bianchi-II universe in the presence of higher spatial curvature terms}

As we have already told, being invented as a tool to solve the problem of the non-renormalizability of quantum gravity, Ho\v{r}ava-Lifshitz approach has opened quite a few opportunities for the cosmological applications 
(for a review see \cite{Hor-cosm}). There are a number of interesting cosmological implications of the Ho\v{r}ava-Lifshitz gravity. 
Let us introduce some general formulae, useful for the Ho\v{r}ava-Lifshitz cosmology.
Following the paper \cite{Hor-cosm}, we  write down the general action for the $z=3$ Ho\v{r}ava-Lifshitz gravity, without thinking of the detailed balance condition and including the lower-derivative terms. This action has the following form:
\begin{eqnarray*}
&&I = I_{kin}+I_{z=3}+I_{z=2}+I_{z=1}+I_{z=0}+I_m,\nonumber \\
&&I_{kin}=\frac{1}{16\pi G}\int Ndt\sqrt{g}d^3x(K^{ij}K_{ij}-\lambda K^2),\nonumber \\
&&I_{z=3}=\int Ndt\sqrt{g}d^3x(c_1\nabla_iR_{jk}\nabla^jR^{jk}+c_2\nabla_iR\nabla^iR\nonumber \\
&&+
c_3 R_i^jR_j^kR_k^i+c_4RR_i^jR_j^i+c_5R^3,\nonumber \\
&&I_{z=2}=\int Ndt\sqrt{g}d^3x(c_6R_i^jR_j^i+c_7R^2),\nonumber \\
&&I_{z=1}=c_8\int Ndt\sqrt{g}d^3x R,\nonumber  \\
&&I_{z=0}=c_9\int Ndt\sqrt{g}d^3x,
\label{hor-cosm-ac}
\end{eqnarray*}
where $I_m$ is the matter action, $N$ is the lapse function, $K$ is the extrinsic curvature tensor and 
$R$ is three-dimensional curvature tensor. 
In what follows, we shall choose the coefficient $\lambda=1$. In this case the kinetic part of the gravitational action, depending on the extrinsic curvature,  coincides with that of the general relativity. We shall consider an empty model, where $I_m = 0$. 
If we choose the spatial geometry as that of the Bianchi-II model, then the potential terms of the first order in spatial curvature are proportional to $\frac{a^2}{b^2c^2}$, the quadratic terms behave as $\frac{a^4}{b^4c^4}$, while the terms of the third order in spatial curvature are proportional to $\frac{a^6}{b^6c^6}$. 

The field equations for the universe where the potential terms of different orders in curvature are present, are rather complicated for the analysis. Thus, we begin with considering  some kind of a ``toy'' model, where only the cubic contributions to the potential are present. 
This model is exactly solvable, and being relatively simple, it, nevertheless,    reveals some interesting qualitative features. 

\subsection{A  model with the only cubic terms in spatial curvature}
In this toy model the equations, analogous to Eqs. (\ref{alpha})--(\ref{gamma}), have the following form:
\begin{equation}
\ddot{\alpha}=-2Le^{4(2\alpha-\beta-\gamma)},
\label{alphaL}
\end{equation}
\begin{equation}
\ddot{\beta}=Le^{4(2\alpha-\beta-\gamma)},
\label{betaL}
\end{equation}
\begin{equation}
\ddot{\gamma}=Le^{4(2\alpha-\beta-\gamma)},
\label{gammaL}
\end{equation}
while Eq. (\ref{00}) is still valid.  Here, the coefficient $L$ represents a linear combination of the coefficients of different cubic contributions to the potential of the model. 
It is convenient to introduce a new variable 
\begin{equation}
x=2\alpha-\beta-\gamma.
\label{x}
\end{equation}
Then, we have
\begin{equation}
\ddot{x}\equiv-6Le^{4x}.
\label{x1}
\end{equation}

Let us first consider the case when $L > 0$. Note that such a choice of the sign of $L$ is most natural, because it coincides with the sign of the potential term 
in the standard general relativity. 
The first integral of Eq. (\ref{x1}) is 
\begin{equation}
\dot{x}^2+3Le^{4x}=H^2.
\label{x2}
\end{equation}
Integrating the equation
\begin{equation}
\dot{x}=\pm\sqrt{H^2-3Le^{4x}},
\label{x3}
\end{equation}
we find  
\begin{equation}
e^{4x}=\frac{4H^2e^{\pm4H\tau}}{3L(e^{\pm4H\tau}+1)^2}.
\label{x4}
\end{equation}
In what follows we shall choose the sign ``plus'' in the exponent.
Substituting the expression (\ref{x4}) into Eq. (\ref{alphaL}) one finds
\begin{equation}
\alpha(\tau)=A\tau +\frac{2}{3}H\tau-\frac16\ln(e^{4H\tau}+1),
\label{alphaL1}
\end{equation}
where $A$ is an integration constant. Analogously,
\begin{equation}
\beta(\tau)=B\tau - \frac{1}{3}H\tau+\frac{1}{12}\ln(e^{4H\tau}+1),
\label{betaL1}
\end{equation}
\begin{equation}
\gamma(\tau)=C\tau - \frac{1}{3}H\tau+\frac{1}{12}\ln(e^{4H\tau}+1).
\label{gammaL1}
\end{equation}
When $\tau \rightarrow \infty$ the expressions (\ref{alphaL1})--(\ref{gammaL1}) behave like 
\begin{eqnarray}
&&\alpha(\tau) = A\tau,\nonumber \\
&&\beta(\tau) = B\tau,\nonumber \\
&&\gamma(\tau) = C\tau.
\label{alphaL2}
\end{eqnarray}
Substituting the expressions (\ref{alphaL2}) into  Eq. (\ref{00}) we find the following relation between the constants $A, B$ and $C$:
\begin{equation}
C=-\frac{AB}{A+B}.
\label{C}
\end{equation}
Now the Kasner indices have the form 
\begin{eqnarray}
&&p_1 = \frac{A}{A+B+C},\nonumber \\
&&p_2 = \frac{B}{A+B+C},\nonumber \\
&&p_3 = \frac{C}{A+B+C}.
\label{KasnerL}
\end{eqnarray}
Using the relation (\ref{C}) we check that the indices (\ref{KasnerL}) satisfy both the Kasner relations (\ref{Kasner}). Hence, it is convenient to use the Khalatnikov-Lifshitz parameter $u$ as 
\begin{equation}
A = - u,\ B = u+1,\ C = u(u+1).
\label{KL-L}
\end{equation}
to come to the standard parametrization of the Kasner indices (\ref{Kasner1}). 

Let us see now what happens when $\tau \rightarrow -\infty$ and we are closer to the singularity.
In this case the scale factor exponents $\alpha, \beta$ and $\gamma$ behave as 
\begin{eqnarray}
&&\alpha(\tau) = A\tau +\frac23H\tau,\nonumber \\
&&\beta(\tau) = B\tau -\frac13H\tau, \nonumber \\
&&\gamma(\tau) = C\tau-\frac13H\tau.
\label{alphaL10}
\end{eqnarray}
Note that the sum 
\begin{equation}
\alpha(\tau)+\beta(\tau)+\gamma(\tau) = (A+B+C)\tau
\label{sum-t}
\end{equation}
has always the same form, the spatial volume of the universe changes according to a simple linear law and the relation between the cosmic time and the logarithmic time is always the same in contrast to the situation which is observed in the General Relativity (see \cite{BKL,Land-Lif} and the preceding section of the present paper).   

Substituting the expressions (\ref{alphaL1}) into Eq. (\ref{00}) we find that 
\begin{equation}
H=B+C-2A= u^2+4u+1.
\label{HL}
\end{equation}
The new Kasner indices are 
\begin{eqnarray}
&&p_1'=\frac{2u^2+5u+2}{3(1+u+u^2)},\nonumber \\
&&p_2'=\frac{-u^2-u+2}{3(1+u+u^2)},\nonumber \\
&&p_3'=\frac{2u^2-u-1}{3(1+u+u^2)}.
\label{KasnerL1}
\end{eqnarray}
We see that as in the case of the General Relativity, there is the change of the signs of the Kasner indices corresponding to the directions 1 and 2. The first Kasner index, which was negative, becomes positive, while the second Kasner index, which was positive, becomes negative.  However, the law of the change of the indices is now more complicated and cannot be expressed 
by means of the shift of the Khalatnikov-Lifshitz parameter $u$. 

Now, as in the preceding section, we consider the situation, when the index $p_1$, describing the evolution along the first axis is positive, while the second index is negative. In this case, 
\begin{equation}
A=u+1,\ B=-u,\ C=u(u+1),\ H = u^2-2u-2.
\label{change}
\end{equation} 
The initial set of the Kasner indices at $\tau \rightarrow \infty$ is now 
\begin{eqnarray}
&&\tilde{p}_1=\frac{u+1}{1+u+u^2},\nonumber \\
&&\tilde{p}_2=-\frac{u}{1+u+u^2},\nonumber \\
&&\tilde{p}_3=\frac{u(u+1)}{1+u+u^2}.
\label{change100}
\end{eqnarray}
When $\tau \rightarrow -\infty$ the Kasner indices change as follows:
\begin{eqnarray}
&&\tilde{p}_1'=\frac{2u^2-u-1}{3(1+u+u^2)},\nonumber \\
&&\tilde{p}_2'=\frac{-u^2-u+2}{3(1+u+u^2)},\nonumber \\
&&\tilde{p}_3'=\frac{2u^2+5u+2}{3(1+u+u^2)}.
\label{change2}
\end{eqnarray}
The new set of the Kasner indices (\ref{change2}) at $\tau \rightarrow -\infty$ has the same signs as those (\ref{change100}) at $\tau \rightarrow \infty$. Thus,  a change of the directions of expansion and contraction in this case is absent in contrast to the case, considered in the preceding section and to the case when the Kasner index, corresponding to the first axis is 
negative (Eqs. (\ref{Kasner1}), (\ref{KasnerL1})). Meanwhile, one can notice a curious relation between the indices (\ref{KasnerL1}) and (\ref{change2}), namely
\begin{equation}
\tilde{p}_1'=p_3',\ \tilde{p}_2'=p_2',\ \tilde{p}_3'=p_1'.
\label{curious}
\end{equation} 

Now, we can investigate what happens if the sign of the cubic in curvature potential terms is different. In other words, we can consider Eqs. (\ref{alphaL}) --(\ref{gammaL}) with $L = - M < 0$, where $M$ is a positive constant. We introduce the variable $x$ as in Eq. (\ref{x}) and find the first integral of Eq. (\ref{x1}), which is now equal to 
\begin{equation*}
\dot{x}^2-3Me^{4x} = const.
\label{first-new}
\end{equation*}
The integration constant  can now be positive, negative or equal to zero. Let us consider all these cases.

First, if
\begin{equation}
\dot{x}^2-3Me^{4x} = 0,
\label{first-new1}
\end{equation}
then 
\begin{eqnarray}
&&\alpha(\tau) = A\tau -\frac16\ln|\tau|,\nonumber \\
&&\beta(\tau)= B\tau +\frac{1}{12}\ln|\tau|,\nonumber \\
&&\gamma(\tau) = C\tau +\frac{1}{12}\ln|\tau|.
\label{zero}
\end{eqnarray}
In this case the behavior of the scale factor at plus and minus infinity is the same and there is no change of the Kasner indices, but some another curious phenomenon arises: at $\tau \rightarrow 0$ the factors $\alpha, \beta$ and $\gamma$ are singular, while their sum, and hence, the spatial volume of the universe remains regular. Thus, we encounter some kind of singularity, which we shall discuss in more detail a little bit later. 

Second, if 
\begin{equation*}
\dot{x}^2-3Me^{4x} = H^2,
\label{first-new2}
\end{equation*}
then the solutions of the equations for the functions $\alpha, \beta$ and $\gamma$ are
\begin{eqnarray}
&&\alpha(\tau)=A\tau +\frac23H\tau-\frac16\ln|e^{4H\tau}-1|,\nonumber \\
&&\beta(\tau) = B\tau -\frac13H\tau+\frac{1}{12}\ln|e^{4H\tau}-1|,\nonumber \\
&&\gamma(\tau)=C\tau -\frac13H\tau+\frac{1}{12}\ln|e^{4H\tau}-1|.
\label{positive}
\end{eqnarray}
If we forget, for a moment, the singularity at $\tau \rightarrow 0$, then one can see that the asymptotical behavior of these functions at $\tau \rightarrow \pm \infty$ coincides with that, considered before for the case with $L > 0$. Hence, the rules of changes of the Kasner indices have the same form. However, at $\tau \rightarrow 0$ we have the same strange singularity which we had already encountered in the case described by the equations  (\ref{first-new1}) and (\ref{zero}). 

Third, let us consider the last possible case, when 
\begin{equation*}
\dot{x}^2-3Me^{4x} = -H^2.
\label{first-new3}
\end{equation*}
Now, the solution for the scale factors has even more exotic form:
\begin{eqnarray}
&&\alpha(\tau) = A\tau -\frac16\ln|\sin 2H\tau|,\nonumber \\
&&\beta(\tau) = B\tau + \frac{1}{12}\ln|\sin 2H\tau|,\nonumber \\
&&\gamma(\tau) = C\tau + \frac{1}{12}\ln|\sin 2H\tau|.
\label{negative}
\end{eqnarray} 
The asymptotic behavior of the functions $\alpha, \beta$ and $\gamma$ at $\tau \rightarrow \pm \infty$
is the same as in the case, described by Eqs. (\ref{first-new1}) and (\ref{zero}). However, in this case we have an infinite number of the singular moments of $\tau$ such that 
\begin{equation*}
\tau = \frac{\pi n}{2H},\ n=0,\pm 1,\pm 2, \cdots,
\label{inf-sing}
\end{equation*}
where we again stumble upon the anisotropic singularities logarithmically divergent at the corresponding moments of time $\tau$. 

\subsection{Big Filament singularity and its properties}

Let us look at the singularity at $\tau \rightarrow 0$ arising in the solutions 
(\ref{zero}) and (\ref{positive}). In the solution (\ref{negative}) the singularity is similar, but it arises in a countable set of time moments. Thus, we shall consider the situation when in the vicinity of the time moment $\tau = 0$, the functions $\alpha, \beta$ and $\gamma$ behave like 
\begin{eqnarray*}
 &&\alpha \sim -\frac16\ln \tau,\nonumber \\
 &&\beta \sim \frac{1}{12}\ln \tau,\nonumber \\
 &&\gamma \sim \frac{1}{12}\ln \tau.
 \label{fila}
 \end{eqnarray*}
 That means that the scale factors $a, b$ and $c$ behave like 
 \begin{eqnarray}
 &&a \sim \tau^{-\frac16},\nonumber \\
 &&b \sim \tau^{\frac{1}{12}},\nonumber \\
 &&c \sim \tau^{\frac{1}{12}},
 \label{fila1}
 \end{eqnarray}
 i.e. the direction 1 is infinitely stretched, while the directions 2 and 3 are infinitely squeezed, but in such a way that the product of the three scale factors, describing the spatial volume of the universe, is finite and regular. We can call this singularity ``Big Filament''. 
 
It is curious also to compare this singularity with that which can arise in the model, where only the linear in the spatial curvature terms are presented, but with a ``wrong sign''. The change of the sign in the right-hand side of Eqs. (\ref{alpha})--(\ref{gamma}) 
implies the following solutions:
 \begin{eqnarray*}
&&\alpha(\tau) =  H\tau -\frac12\ln(e^{4H\tau}-1),\nonumber \\
&&\beta(\tau) = B\tau - H\tau + \frac12\ln(e^{4H\tau}-1),\nonumber \\
&&\gamma(\tau)=C\tau - H\tau + \frac12\ln(e^{4H\tau}-1).
\label{alpha3wrong}
\end{eqnarray*} 
Obviously, at $\tau \rightarrow 0$ we also encounter singularity and the scale factors behave as 
\begin{eqnarray*}
&&a(\tau) \sim \frac{1}{\sqrt{\tau}},\nonumber \\
&&b(\tau) \sim \sqrt{\tau}, \nonumber \\
&&c(\tau) \sim \sqrt{\tau}.
\label{fila2}
\end{eqnarray*}
Also in this case the scale factor $a(\tau)$ becomes infinitely big, while the factors $b(\tau)$ and $c(\tau)$ are infinitely small. However, in contrast to the situation described by Eq. (\ref{fila1}), the spatial volume $abc \sim \sqrt{\tau}$ and becomes infinitely small. Thus, in this case the Big Filament singularity becomes stronger.  One can also consider the model where only the quadratic in the spatial curvature terms are present. Choosing in a proper way the signs of these terms, one can come to the solutions which at $\tau \rightarrow 0$ behave like 
\begin{eqnarray*}
&&a(\tau) \sim \tau^{-\frac{5}{21}},\nonumber \\
&&b(\tau) \sim \tau^{\frac17},\nonumber \\
&&c(\tau) \sim \tau^{\frac17}.
\label{fila3}
\end{eqnarray*}
Also in this case we have a Big Filament type singularity, where one of the spatial dimensions tends to infinity, while two other spatial dimensions and the spatial volume tend to zero. Thus, we see that only the cubic in spatial curvature term can produce 
a particular, relatively mild, anisotropic singularity, described by Eqs. (\ref{fila1}). It looks like the universe can cross this singularity, even if it is impossible for the extended objects to cross it without being crushed.

\subsection{The exact solutions for a  model with the only potential terms, coming from the quadratic in spatial curvature structures}
At the end of this section we would like also to write down the exact solutions for the model, where in the equations of motion  only the quadratic terms are present:
\begin{eqnarray}
&&\ddot{\alpha}=-5Ke^{2(3\alpha-\beta-\gamma)}, \nonumber \\
&&\ddot{\beta}=3Ke^{2(3\alpha-\beta-\gamma)}, \nonumber \\
&&\ddot{\gamma}=3Ke^{2(3\alpha-\beta-\gamma)}.
\label{quadratic}
\end{eqnarray}
We shall consider only the case, when $K > 0$. Then, proceeding just like in the section 2 and in the subsection 3.1 we can obtain the following general solution:
\begin{eqnarray*}
&&\alpha(\tau) = A\tau + \frac{10}{21}H\tau-\frac{5}{21}\ln(e^{2H\tau}+1),\nonumber \\
&&\beta(\tau)=B\tau-\frac27H\tau+\frac17\ln(e^{2H\tau}+1),\nonumber \\
&&\gamma(\tau)=C\tau-\frac27H\tau+\frac17\ln(e^{2H\tau}+1).
\label{quadratic1}
\end{eqnarray*}
If far away from the singularity at $\tau \rightarrow \infty$ one has a Kasner regime with $p_1<p_2<p_3$ and with the standard parametrisation (\ref{Kasner1}), then approaching to the singularity at $\tau \rightarrow -\infty$ we obtain the Kasner regime with the indices 
\begin{eqnarray}
&&p_1' = \frac{10u^2+29u+10}{19u^2+11u+19},\nonumber \\
&&p_2'= \frac{-6u^2-9u+15}{19u^2+11u+19},\nonumber \\
&&p_3'=\frac{15u^2-9u-6}{19u^2+11u+19}.
\label{quadratic2}
\end{eqnarray} 
These formulas will be used in the next two sections. 

\section{The mixmaster universe in the Hamiltonian formalism}
In this section  we compare our results with those, obtained in paper \cite{Bakas}, where the Hamiltonian formalism was used. It is convenient for us to use the notations close to those used in  the paper \cite{Bakas}. 

The metric of a homogeneous Bianchi universe can be written as 
\begin{equation*}
ds^2=N^2(t)dt^2-a^2(t)\sigma_1^2-b^2(t)\sigma_2^2-c^2(t)\sigma_3^2, 
\label{Bakas}
\end{equation*} 
where $N$ is a lapse function. 
It is convenient to parametrize the scale factors $a,b$ and $c$ as follows:
\begin{eqnarray*}
&&a=e^{\Omega+\frac{\beta_+}{2}+\frac{\sqrt{3}}{2}\beta_-},\nonumber \\
&&b=e^{\Omega+\frac{\beta_+}{2}-\frac{\sqrt{3}}{2}\beta_-},\nonumber \\
&&c=e^{\Omega-\beta_+}.
\label{Bakas1}
\end{eqnarray*}
 If the lapse function $N$ is chosen to be proportional to the spatial volume of the universe, then the role of the time parameter
 plays the logarithmic time $\tau$, used in the preceding sections.
Introducing the conjugate momenta, one arrives to the gauge-fixed Hamiltonian  
\begin{equation}
H = \frac{1}{2}\left(p_+^2+p_-^2-\frac14p_{\Omega}^2\right)+V(\beta_+,\beta_-,\Omega).
\label{Bak-Ham}
\end{equation}
If we consider a Bianchi-II universe and General Relativity then the potential $V$ has the form
\begin{equation}
V=V_0e^{4(\Omega-\beta_+)}.
\label{Bak-pot}
\end{equation}
In the case of the Bianchi-IX universe the structure of the potential is more complicated \cite{Misner,Bakas} and the dynamics of the system can be represented as a motion of a ball in the billiard with moving walls. The billiard is two-dimensional $(\beta_+,\beta_-)$ and the parameter $\Omega$ plays the role of a time variable. The dependence of the potential on the $\Omega$ 
means that the walls of the billiard are moving. The expression (\ref{Bak-pot}) arises as an asymptotical form of the full potential in the Bianchi-IX model, when we consider the bounce of the ball from one of the walls, or, in other words, the change of the Kasner regime. Indeed, when the ball is far away from the walls, and we can neglect the potential, it is moving freely 
and the corresponding universe is a Kasner one.  

Now, we write down the relations connecting the Kasner indices with the conjugate momenta \cite{Bakas}. They look as follows:
\begin{eqnarray}
&&p_1=\frac{p_{\Omega}+4p_+}{3p_{\Omega}},\nonumber\\
&&p_2=\frac{p_{\Omega}-2p_+-2\sqrt{3}p_-}{3p_{\Omega}},\nonumber\\
&&p_3=\frac{p_{\Omega}-2p_++2\sqrt{3}p_-}{3p_{\Omega}}.
\label{Bak-Kas}
\end{eqnarray}
Inversely,
\begin{eqnarray}
&&\frac{p_+}{p_{\Omega}}=\frac{3p_1-1}{4},\nonumber \\
&&\frac{p_-}{p_{\Omega}}=\frac{\sqrt{3}(p_3-p_2)}{4}.
\label{Bak-Kas1}
\end{eqnarray}
Note that the Kasner indices (\ref{Bak-Kas}) satisfy the condition $p_1^2+p_2^2+p_3^2=1$ due to the Hamiltonian constraint 
$$
p_{\Omega}^2=4(p_+^2+p_-^2),
$$
which is valid far away from the wall, where the potential is negligible. 

Other useful formulae are given by the Misner parametrization of the Kasner indices \cite{Misner}:
\begin{eqnarray}
&&p_1=-\frac{(s-3)(s+3)}{3(s^2+3)},\nonumber \\
&&p_2=\frac{2s(s-3)}{3(s^2+3)},\nonumber \\
&&p_3=\frac{2s(s+3)}{3(s^2+3)}.
\label{Misner-Kasner}
\end{eqnarray}

Now, coming back to the Hamiltonian (\ref{Bak-Ham}) with the potential (\ref{Bak-pot}), we can make a change of variables, eliminating the dependence of the potential on the spatial volume $\Omega$:
\begin{eqnarray}
&&\bar{\beta}_+=\frac{1}{\sqrt{3}}(\beta_+-\Omega),\nonumber \\
&&\bar{\Omega}=\frac{1}{2\sqrt{3}}(4\Omega-\beta_+).
\label{new-var}
\end{eqnarray}
The corresponding conjugate momenta are 
\begin{eqnarray}
&&\bar{p}_+=\frac{1}{\sqrt{3}}(4p_++p_{\Omega}),\nonumber \\
&&\bar{p}_{\Omega}=\frac{2}{\sqrt{3}}(p_++p_{\Omega}).
\label{new-var1}
\end{eqnarray}
Now, the Hamiltonian looks as 
\begin{equation}
H=\frac12\left(\frac14\bar{p}_+^2+p_-^2-\frac14\bar{p}_{\Omega}^2\right)+V_0e^{-4\sqrt{3}\bar{\beta}_+}.
\label{Bakas-Ham1}
\end{equation}
The conjugate momenta $p_-$ and $\bar{p}_{\Omega}$ are constant because the potential in Eq. (\ref{Bakas-Ham1}) does not depend on $\beta_-$ and $\bar{\Omega}$.   
Hence, one can state that also the relation 
\begin{equation}
\frac{\bar{p}_{\Omega}}{p_-}=\frac{2}{\sqrt{3}}\frac{\frac{p_+}{p_{\Omega}}+1}{\frac{p_-}{p_{\Omega}}} = const.
\label{const}
\end{equation}
This relation connect the values of the momenta at two Kasner regimes. Substituting into Eq. (\ref{const}) the relations 
(\ref{Bak-Kas1}) and (\ref{Misner-Kasner}) we find that
\begin{equation}
\frac{s}{3}+\frac{3}{s} = const.
\label{const1}
\end{equation}
The formula (\ref{const1}) signifies that the only change of the variable $s$, leaving the left-hand side of the relation intact is 
\begin{equation}
s \rightarrow \frac{9}{s}.
\label{change}
\end{equation}
It is well known that the change of the Kasner regime described in terms of the Misner parameter $s$ by the formula (\ref{change}) coincides with those, described in terms of the Khalatnikov-Lifshitz parameter in the Section 2. We shall not dwell on this fact and will go directly to the description of the change of the Kasner regime for the models driven by potential coming from the terms cubic or quadratic in spatial curvature. 

In the case when only the cubic term is present the potential in the Hamiltonian has the form
\begin{equation}
V = V_1e^{-12\beta_+}.
\label{pot-cub}
\end{equation}
The potential (\ref{pot-cub}) does not depend on the volume $\Omega$ and the corresponding conjugate momentum $p_{\Omega}$ is constant, as well as the relation $p_/p_{\Omega}$. Using the formulas (\ref{Bak-Kas1}) and (\ref{Misner-Kasner}) we find that \cite{Bakas}
\begin{equation}
\frac{s}{\sqrt{3}}+\frac{\sqrt{3}}{s} = const.
\label{const10}
\end{equation}
The corresponding change of the variable $s$ is 
\begin{equation}
s \rightarrow \frac{3}{s}.
\label{change1}
\end{equation}
Making this change in the formulas (\ref{Misner-Kasner}) we obtain
\begin{eqnarray}
&&p_1'=\frac{s^2-1}{s^2+3},\nonumber \\
&&p_2'=-\frac{2(s-1)}{s^2+3},\nonumber \\
&&p_3'=\frac{2(s+1)}{s^2+3}.
\label{Misner-Kasner1}
\end{eqnarray}
To check that these formula coincide with the formulas (\ref{KasnerL1}), obtained in the section 3.1 by the direct resolution of the equations of motion in a Bianchi-II universe, it is convenient to rewrite the new Kasner indices as  combinations of the old Kasner indices:
\begin{eqnarray}
&&p_1'=\frac13(2p_3+2p_2-p_1),\nonumber \\
&&p_2'=\frac13(-p_3+2p_2+2p_1),\nonumber \\
&&p_3'=\frac13(2p_3-p_2+2p_1).
\label{Misner-Kasner2}
\end{eqnarray}
Substituting expressions (\ref{Misner-Kasner}) into Eqs. (\ref{Misner-Kasner2}) we obtain the expressions 
(\ref{Misner-Kasner1}), as it should be.

The quadratic in curvature terms give the following structure of the potential for the Bianchi-II model:
\begin{equation}
V=V_2e^{2\Omega-8\beta_+}.
\label{quad-pot}
\end{equation}
Now it is convenient to introduce the following new variables:
\begin{eqnarray}
&&\bar{\beta}_+=\frac{4}{3\sqrt{7}}\left(\beta_+-\frac14\Omega\right),\nonumber \\
&&\bar{\Omega}=\frac{1}{6\sqrt{7}}(16\Omega-\beta_+).
\label{quad-pot-new}
\end{eqnarray}
The new momenta are
\begin{eqnarray}
&&\bar{p}_{+}=\frac{127}{24\sqrt{7}}p_++\frac{1}{6\sqrt{7}}p_{\Omega},\nonumber \\
&&\bar{p}_{\Omega}=\frac{1}{3\sqrt{7}}(4p_{\Omega}+p_+).
\label{quad-pot-new1}
\end{eqnarray}
The Hamiltonian looks now as 
\begin{equation}
H=\frac12\left(\frac14\bar{p}_+^2+p_-^2-\frac14\bar{p}_{\Omega}^2\right)+V_2e^{-3\sqrt{7}\bar{\beta}_+}.
\label{Bakas-Ham2}
\end{equation}
In this case the relation $\bar{p}_{\Omega}/p_-$ is again constant and using the formulas (\ref{quad-pot-new1}) and (\ref{Bak-Kas1}) we obtain
$$
\frac{p_1+5}{p_3-p_2} = const,
$$ 
which in turn, on implying the relations (\ref{Misner-Kasner}) gives
\begin{equation*}
\sqrt{\frac{7}{27}}s+\sqrt{\frac{27}{7}}\frac1s = const.
\label{relation3}
\end{equation*}
The last relation implies the change of variables 
\begin{equation*}
s \rightarrow \frac{27}{7}\frac1s,
\label{change20}
\end{equation*}
describing the change of the Kasner regime. 
Making this change of the variables in the formulas (\ref{Misner-Kasner}), we obtain
\begin{eqnarray}
&&p_1'=\frac{49s^2-81}{49s^2+243},\nonumber \\
&&p_2'=\frac{18(9-7s)}{49s^2+243},\nonumber \\
&&p_3'=\frac{18(9+7s)}{49s^2+243}.
\label{change3}
\end{eqnarray}
 To compare the formulas (\ref{change3}) with those obtained in the end of the third section (\ref{quadratic2}), it is convenient to rewrite the formulas (\ref{quadratic2}) in the following form:
 \begin{eqnarray}
 &&p_1'=\frac{10p_3+10p_2-9p_1}{19p_3+19p_2+27p_1},\nonumber \\
 &&p_2'=\frac{-6p_3+15p_2+18p_1}{19p_3+19p_2+27p_1},\nonumber \\
 &&p_3'=\frac{15p_3-6p_2+18p_1}{19p_3+19p_2+27p_1}.
 \label{change4}
 \end{eqnarray}
Substituting into the formulas (\ref{change4}) the expressions (\ref{Misner-Kasner}) we again reproduce the formulas (\ref{change3}). Thus, also in this case 
the Hamiltonian formalism and the solution of equations for the Bianchi-II universe give the same results.

\section{Evolution of the Mixmaster universe in the Ho\v{r}ava-Lifshitz inspired models: BKL approach}

In this section, using the results obtained in the Section 3, we shall describe the evolution of the Mixmaster universe, using the BKL approach \cite{BKL} 
and the Khalatnikov-Lifshitz parameter $u$ \cite{Khal-Lif}. Before the consideration of the Ho\v{r}ava-Lifshitz models, we reproduce the laws of the change of the Kasner epochs and eras in General Relativity. In the Section 2, considering the exact solution for the Bianchi-II model and starting with the asymptotic Kasner regime  with $p_1 \leq p_2 \leq p_3$ and the standard parametrisation (\ref{Kasner1}) we have obtained that the final (closer to the cosmological singularity) Kasner regime is characterised by the Kasner indices (\ref{Kasner2}). Let us look at these formulae more attentively. 
We shall try to reproduce the known formulae for the changes of the Kasner epochs and eras \cite{BKL} with more detail then it was usually done to simplify the understanding of the corresponding transformations in theories which differ from general relativity. 
We shall begin with the analysis of the formula, describing the new value of the second Kasner index:
\begin{equation}
p_2'=\frac{1-u}{1-u+u^2}.
\label{p2}
\end{equation}
Let us remember, that we always use the values of the parameter $u$ such that $1 \leq u < \infty$. Thus, the expression (\ref{p2}) is negative. Now we can 
rewrite the expressions for the new Kasner indices, using again the Khalatnikov-Lifshitz parameter, but this parameter should change its value in such a way to represent the index $p_2'$ as a negative Kasner index $p_1$ with this new value of the Khalatnikov-Lifshitz parameter . Namely, we should solve the following equation:
\begin{equation}
\frac{1-u}{1-u+u^2} = \frac{-u'}{1+u'+u'^2}.
\label{p21}
\end{equation}
 Being quadratic, the above equation has two  solutions:
 \begin{equation}
 u'=u-1,
 \label{p220}
 \end{equation}
 \begin{equation}
 u'=\frac{1}{u-1}.
 \label{p22}
 \end{equation}
 Inversely,
 \begin{equation}
 u=u'+1, 
 \label{p230}
 \end{equation}
 \begin{equation}
 u=\frac{1+u'}{u'}.
 \label{p23}
 \end{equation}
 Which of two solutions (\ref{p220}) or (\ref{p22}) should we use?
 If $u>2$ then $u'=u-1$ is also greater than $1$ and we can choose the solution (\ref{p220}).
 Substituting the expression (\ref{p230}) into Eq. (\ref{Kasner2}) one obtains
 \begin{eqnarray}
 &&p_1'=p_2(u-1),\nonumber \\
 &&p_2'=p_1(u-1),\nonumber \\
 &&p_3'=p_3(u-1).
 \label{p24}
 \end{eqnarray}
We see that the axes $1$ and $2$ exchange their roles, because the new Kasner index $p_1'$ becomes positive 
while the new Kasner index $p_2'$ becomes negative. Such a change is called the change of the Kasner epoch \cite{BKL}.
If, instead, $u < 2$ then $u'=u-1$ is less than $1$. In this case we should choose the solution (\ref{p22}). Then substituting 
(\ref{p23}) into Eq. (\ref{Kasner2}) we obtain 
\begin{eqnarray}
&&p_1'=p_3\left(\frac{1}{u-1}\right),\nonumber \\ 
&&p_2'=p_1\left(\frac{1}{u-1}\right),\nonumber \\
&&p_3'=p_2\left(\frac{1}{u-1}\right).
\label{p25}
\end{eqnarray}
Such a change of the Kasner regime is called the change of a Kasner era \cite{BKL}. 
  
In the case of the Bianchi-IX universe the transition from one Kasner regime to another is not the end of the evolution in contrast to the case of the Bianchi-II universe. The final Kasner regime is indeed the initial regime for the new transition, which will be driven by another term in spatial curvature, which becomes dominant. 
A Kasner era can possess an arbitrary number of Kasner epochs - it is equal to the integer part of the number $u$, which characterises the beginning of the Kasner era. The transition  from one Kasner era to another is described by the so called Gauss transformation, which is responsible for the stochasticity of this evolution \cite{chaos}.    
Let us also add that in the case of the change of the Kasner epoch the new values of the Kasner indices satisfy the inequality 
\begin{equation*}
p_2' \leq p_1' \leq p_3',
\label{epoch}
\end{equation*}
while after the change of the Kasner era 
we have 
\begin{equation*}
p_2' \leq p_3' \leq p_1'.
\label{era}
\end{equation*}
Note that if the parameter $u$ has a rational value, then after a finite number of the changes of the Kasner epochs and eras, one arrives to the value $u=1$, which gives the transition to the regime with $u=0$, i.e. the regime with two Kasner indices equal to zero, while the third one is equal to one. That means that  the metric has (up to the permutation of the axes) the form 
\begin{equation}
ds^2=dt^2-t^2dx^2-dy^2-dz^2.
\label{Mink}
\end{equation}   
This is nothing but the Minkowski spacetime,  represented in terms of  non-standard coordinates \cite{Land-Lif}. Such evolutions, ended in the Minkowski spacetime constitute a set of  measure zero.    

It is clear from the formulae given above that the Kasner eras with  arbitrary numbers of the Kasner epochs can exist. Moreover, one can always find such values of the parameter $u$ which provide us with a desirable succession of the Kasner eras. A convenient tool relating the length of the successive Kasner eras
with the parameter $u$ is the continuous fraction expansion of this parameter \cite{BKL,Land-Lif}, but we shall not dwell on this topic. Instead, we would like to say some words about the topological entropy and symbolic dynamics, which are very useful and relatively simple tools for the description of  chaotic systems \cite{topol,topol1,topol2}. 
Let us first say that the phase space of the Bianchi - IX model in the General Relativity contains an infinite set of unstable periodic trajectories. The term ``periodic'' means that after some finite number of the changes of the Kasner epochs and eras the universe again finds itself in the Kasner regime with the same   
 values of the Kasner indices, which it had at the beginning of the cycle. In other words, after some finite series of transformations, the parameter $u$ acquires its initial value.  Every periodic trajectory can be characterised symbolically as a succession of two letters, say $A$ and $B$, corresponding to the changes of the epochs and of the eras, respectively \cite{topol}.  Obviously, all the possible successions of these letters are acceptable. Thus, the number of the trajectories of the length $k$ is $N(k) = 2^k$ \cite{topol}.  The topological entropy is defined as 
 \begin{equation}
 H_T=\lim_{k\rightarrow\infty}\frac{1}{k}\ln N(k). 
 \label{topol}
 \end{equation}
The positivity of the topological entropy means that the dynamics of the system under consideration is chaotic. Substituting $N(k) = 2^k$ into the definition (\ref{topol}) we see that $H_T=\ln 2$. Hence, we have another confirmation of the fact that the dynamics of the oscillating approach to the singularity in the Bianchi-IX universe in General Relativity is chaotic. In terms of the Hamiltonian dynamics (see the preceding section and the references therein) this corresponds
to the chaotic reflections of the ball from the moving walls of the billiard. 

Now, using the results of the  subsection 3.1 we shall describe what happens in the mixmaster model, where only the strongest (cubic) in spatial curvature terms
are present. We have seen that the consideration of the Bianchi-II model governed by the equations (\ref{alphaL})--(\ref{gammaL}) implies the change of the Kasner indices, given by the formulas (\ref{KasnerL1}). The second Kasner index becomes negative and we can try a new value of the Khalatnikov-Lifshitz parameter $u$, resolving the equation
\begin{equation}
p_2'=\frac{-u^2-u+2}{3(1+u+u^2)}=\frac{-u'}{1+u'+u'^2}.
\label{new-u}
\end{equation}  
Its solutions are 
\begin{eqnarray}
&&u'_1=\frac{u+2}{u-1},\nonumber \\
&&u'_2=\frac{u-1}{u+2}.
\label{new-u1}
\end{eqnarray}
Obviously, independently on the value of $u>1$, $u'_1 > 1$ and we should choose 
\begin{equation}
u'=T(u)=\frac{u+2}{u-1},
\label{new-u10}
\end{equation}
where the letter $T$ symbolises the operation of the transformation of the parameter $u$.
Then, the relations (\ref{KasnerL1}) can be represented as 
\begin{eqnarray}
&&p_1'=p_3\left(\frac{u+2}{u-1}\right),\nonumber \\
&&p_2'=p_1\left(\frac{u+2}{u-1}\right),\nonumber \\
&&p_3'=p_2\left(\frac{u+2}{u-1}\right)
\label{new-u2}
\end{eqnarray}
and 
\begin{equation}
p_2' \leq p_3' \leq p_1'.
\label{new-u3}
\end{equation}
That means that a change of the Kasner era has taken place. Using this result we can say that in the mixmaster Bianchi-IX universe all the Kasner eras contain one and only one Kasner epoch and only one type of the change of the Kasner regime does exist. It means that the regime of the approaching of the singularity is oscillating, but not chaotic. Moreover, applying the transformation (\ref{new-u10}) two times we come back to the initial value of $u$, namely
\begin{equation}
T(T(u)) =T\left(\frac{u+2}{u-1}\right)= u.
\label{new-u4}
\end{equation}
That means that all the evolutions in this case are periodic. We can describe them in more detail.
If at the beginning we have 
$$p^{(0)}_1\leq p^{(0)}_2 \leq p^{(0)}_3$$
and the Khalatnikov - Lifshitz parameter has a certain value $u$ then after the first change of the Kasner regime we have 
$$p^{1}_2 \leq p^{(1)}_3 \leq p^{(1)}_1$$ 
with the parameter, which has the value $u'=\frac{u+2}{u-1}$. After the second transition, one has 
$$p^{(2)}_3 \leq p^{(2)}_1\leq p^{(2)}_2,$$ 
while the parameter again has the value $u$. After the third change of the regime, one obtains 
$$p^{(0)}_1\leq p^{(0)}_2 \leq p^{(0)}_3,$$ 
i.e. the order of the Kasner indices is the same as it was at the beginning of the cycle, but the value of the parameter is $u'$. Continuing 
the changes, we see that after the six transitions, we reproduce both the order of the Kasner indices and the initial value of the parameter $u$. Thus, the periodicity,
described in paper \cite{Bakas} in terms of the Hamiltonian formalism, has a very simple form in the BKL formalism. One can note a couple of other interesting facts. First, it is impossible starting from $u>1$ arrive to $u=0$, that means that the Minkowski regime is unaccessible in this model in contrast to the mixmaster dynamics in General Relativity. This fact was described in \cite{Bakas} in the billiard terms. Second, resolving equation $T(u) = u$, which gives 
$u_0 = \sqrt{3}+1$, we see that in this case the value of the parameter $u$ never changes. 

Now, let us suppose that only quadratic in spatial curvature terms are present in the action and in the equations of motion. That means that the dynamics of the Bianchi-II universe is governed by Eqs.   (\ref{quadratic}) and the change of the Kasner regime is described by the formulas (\ref{quadratic2}). 
Thus, we can find the law of the change of the parameter $u$, resolving the following equation:
\begin{equation}
p_2'=\frac{-6u^2-9u+15}{19u^2+11u+19}=\frac{-u'}{1+u'+u'^2}.
\label{quad10}
\end{equation}
The two solutions are 
\begin{equation}
u'_1 = T_1(u)=\frac{3u-3}{2u+5},
\label{quad20}
\end{equation}
\begin{equation}
u'_2=T_2(u)=\frac{2u+5}{3u-3}.
\label{quad30}
\end{equation}
Now, analysing an inequality 
\begin{equation}
\frac{3u-3}{2u+5} > 1,
\label{quad40}
\end{equation}
we see that if $u > 8$ then 
\begin{equation}
1 < u'_1 < \frac32.
\label{quad50}
\end{equation}
Hence, the new Kasner indices can be represented as 
\begin{eqnarray}
&&p'_1=p_2\left(\frac{3u-3}{2u+5}\right),\nonumber \\
&&p'_2=p_1\left(\frac{3u-3}{2u+5}\right),\nonumber \\
&&p'_3=p_3\left(\frac{3u-3}{2u+5}\right),
\label{quad60}
\end{eqnarray}     
i.e. we have a change of the Kasner epoch. 
If $u < 8$, then $u'_2 > 1$, the new Kasner indices are expressed as 
\begin{eqnarray}
&&p'_1 = p_3\left(\frac{2u+5}{3u-3}\right),\nonumber \\
&&p'_2 =p_1\left(\frac{2u+5}{3u-3}\right),\nonumber \\
&&p'_3=p_2\left(\frac{2u+5}{3u-3}\right)
\label{quad70}
\end{eqnarray}
and we see a change of a Kasner era. Note, that after a change of a Kasner epoch, the value of the parameter $u$ is $u'_1 < \frac32 < 8$, that means that 
after a change of a Kasner epoch, we inevitably observe a change of a Kasner era. Thus, any Kasner era can contain only $1$ or $2$ Kasner epochs.  

Now, let us suppose that we are at the beginning of a Kasner era, which contains two epochs and, hence, it is characterized 
by the value of the parameter $u > 8$. Can this era be preceded by another Kasner era, containing two epochs? Let us suppose that the preceding era  is characterized by  $\tilde{u} > 8$. Then, the values of the parameters of two eras are connected by the relation
\begin{equation}
T_2(T_1(\tilde{u})) = \frac{16\tilde{u}+19}{3\tilde{u}-24}=u.
\label{era}
\end{equation}
Resolving this relation with respect to $\tilde{u}$, we find 
\begin{equation}
\tilde{u}=\frac{24u+19}{3u-16}.
\label{era1}
\end{equation}
That means that for every value of $u>8$ exists a value $\tilde{u} > 8$ such that the relation (\ref{era}) is satisfied. Hence, every Kasner era, containing two Kasner epochs can be preceded by another Kasner era containing two epochs.   
 Analyzing Eqs. (\ref{era}) and (\ref{era1}), we can see also that  if $\tilde{u} > 26\frac38$, then  a Kasner era, containing two Kasner epochs can be succeeded by a Kasner era containing only one Kasner epoch. Now, let us see what happens, after a Kasner era, containing one Kasner epoch and characterized by some value of the parameter $\tilde{u} < 8$.  In this case 
\begin{equation}
T_2(\tilde{u}) = \frac{2\tilde{u}+5}{3\tilde{u}-3} = u.
\label{era2}
\end{equation}
One sees that if $\tilde{u} < 1\frac{7}{22}$ then after this era follows an era with $u > 8$ and two Kasner epochs, while after
$\tilde{u} > 1\frac{7}{22}$ one has a Kasner era with one epoch. 

All written above implies that one can have arbitrary sequences of the Kasner eras with one or two epochs. In terms of the symbolic dynamics, used for  the description of the mixmaster dynamics in  General Relativity and presented at the beginning of this section,  that means that 
one can have arbitrary  successions of the letters $A$ and $B$, with the only prohibition rule: one cannot have more than one letter $A$ staying together, which corresponds to the fact that a Kasner era cannot contain more than two epochs. 
Curiously, the similar symbolic dynamics was considered in paper \cite{topol1}, where the closed Friedmann model at the presence of the massive scalar field was analyzed (see also paper \cite{topol2} with more complicated potentials).  
There two letters corresponded to the oscillations of the scalar fields and to the bounces of the scale factor. Two bounces of the scale factor could not follow each other without oscillations of the scalar field between them. Thus, the number of the periodic trajectories in the model in \cite{topol1} coincided with that which we study now. It was shown that the numbers $N(k)$ of the successions of $k$ letters $A$ and $B$, satisfying the prohibition rule, mentioned above, constituted the Fibonacci succession,
satisfying the recurrent relation 
$$
N(k+2)=N(k+1)+N(k).
$$
Using the general formula, connecting the Fibonacci numbers with the golden ratio, one can calculate the topological entropy
(\ref{topol}), which in this case is equal to 
$$
H_T = \ln\left(\frac{\sqrt{5}+1}{2}\right) > 0.
$$
Thus, in the case of the model driven by the quadratic in spatial curvature terms, we have a chaotic oscillating approach to the singularity. This corresponds to the reflections of the ball in the billiard with moving walls, described in \cite{Bakas} and in the section 4 of the present paper. 

One can mention another curious fact. If one has the Kasner era with $u = 8$, then the following Kasner era 
will have   $u=1$, which implies the subsequent transition to the Minkowski regime (\ref{Mink}). The value of $u=8$, in turn, can be obtained from $u=26\frac38$ or from $u=1\frac{7}{22}$ etc. Thus, in this case some exceptional evolutions ending in the Minkowski spacetime are present, as it should be in the model of the billiard with the moving walls \cite{Bakas}.

Concluding this section, we would like to say that in our opinion the fact that the chaoticity is present in the model, driven only by quadratic in spatial curvature terms, makes  plausible the following hypothesis.  In more comlicated  Ho\v{r}ava-Lifshitz models, where the cubic terms are present and the quadratic terms are subleading, the presence of the latter can transform the non-chaotic dynamics of the model, where only the leading terms are present into a chaotic one. We think that this hypothesis is worth of further studies.   

\section{Concluding remarks}
Let us recapitulate briefly what was done in the present paper. 
We have considered some Ho\v{r}ava-Lifshitz gravity inspired cosmological models, using the Bianchi-II homogeneous anisotropic spatial geometry. First, we have written down the exact solution for the Bianchi-II universe in the framework of the standard General Relativity. As is well known there is an essential asymmetry between three scale factors in the Bianchi-II cosmology. Namely, 
all the components of the spatial curvature tensor are proportional to the fraction $\frac{a^2}{b^2c^2}$.  If far away from the cosmological singularity the scale factor $a$ is characterised by a negative Kasner index, then coming closer to the singularity 
the axes 1 and 2 exchange their role and the law of transformation of the Kasner indices is described by the shift of the Lifshitz-Khalatnikov parameter $u \rightarrow u-1$ \cite{BKL,Land-Lif}. If, instead, at the beginning, the Kasner index for the scale factor $a$ is positive, the factors $a$ and $b$ again change their roles in the process of the evolution, but the law of the transformation of the Kasner indices is described by the shift $u \rightarrow u+1$.  

Turning to the Ho\v{r}ava-Lifshitz inspired cosmological models, we first considered a toy model, where only cubic in spatial curvature terms were present in the action. The dynamical equations in such a model are also exactly integrable. We presented their solutions 
in an explicit form. In the case when the  Kasner index of the scale factor $a$ is negative, the scale factors $a$ and $b$ change their roles, but the law of the transformation of the Kasner indices (\ref{KasnerL1}) is more complicated than that in the General Relativity (\ref{Kasner2}). If the initial Kasner index for the scale factor $a$ is positive, then in contrast to the General Relativity
case (\ref{Kasner-new}), the Kasner indices conserve their signs (\ref{change2}). 

Our toy model reveals also another curious feature: if we choose the sign in front of the cubic term in a certain way, a new cosmological singularity appears at $\tau \rightarrow 0$. At this point, the scale factor $a$ becomes infinitely large, other two scale factors are infinitely small, while their product (spatial volume) is finite (\ref{fila}). We call this singularity ``Big Filament''. 
This singularity looks softer than usual types of singularities, although the extended objects cannot  cross it.


Concluding, let us remember that the Bianchi-IX cosmological model is much more interesting for the study of the approach to the cosmological singularity \cite{BKL}, because in this model three types of the curvature terms $\frac{a^2}{b^2c^2}, \frac{b^2}{a^2c^2}$ and $\frac{c^2}{a^2b^2}$ are present. Hence, the initial regime of the Kasner-like universe defines which of these three terms will be dominating. The considerations made in the present paper sustain a belief that the combination of the standard and Ho\v{r}ava-Lifshitz terms in the equations of motion of such a general geometry will produce the chaotic behaviour like that, observed in the General Relativity \cite{chaos}, but with more complicated rules of evolution.
It is interesting to note that the occurrence of chaos in the toy model characterized by the quadratic terms in the spatial curvature is rather surprising. Indeed, in the quadratic $f(R)$ theories of gravity like the Starobinsky model \cite{Star.Infl.}, it was shown that the stochastic oscillatory approach to the singularity is suppressed \cite{BarrowCotsakis1,BarrowSiro,MorrMontCapozz} by the  Ricci scalar squared term. 
Thus, the predictions of the modified gravity theories, with unbroken Lorentz invariance, which attract now a growing attention of researchers, differ from those coming from Ho\v{r}ava-Lifshitz gravity. 
 We believe that this problem like other questions, studied here, deserves further studies.  
     
\ack
The work of  A.K. was partially supported by the RFBR grant 17-02-01008.

\end{document}